\documentclass{article}

\usepackage{PRIMEarxiv}

\usepackage[utf8]{inputenc} 
\usepackage[T1]{fontenc}    
\usepackage{url}            
\usepackage{booktabs}       
\usepackage{amsfonts}       
\usepackage{nicefrac}       
\usepackage{microtype}      
\usepackage{lipsum}
\usepackage{graphicx}       
\usepackage{rotating}
\usepackage{gensymb}
\usepackage{siunitx}

\graphicspath{{media/}}     

\title{Printing on particles: combining two-photon nanolithography and capillary assembly to fabricate multi-material microstructures
}

\author{
  Steven van Kesteren\thanks{equal contributions}, Xueting Shen$^{*}$, Michele Aldeghi, Lucio Isa \\
  Laboratory for Soft Materials and Interfaces \\
  Department of Materials, ETH Zurich, Zurich, Switzerland \\
  \texttt{lucio.isa@mat.ethz.ch} \\
}

\begin{document}
\maketitle

\begin{abstract}

Additive manufacturing at the micro- and nanoscale has seen a recent upsurge to suit the increasing demand for more elaborate structures. However, the integration and precise placement of multiple distinct materials at small scales remain a challenge. To this end, we combine here the directed capillary assembly of colloidal particles and two-photon direct laser writing (DLW) to realize a new class of multi-material microstructures. We use DLW both to fabricate 3D micro-templates to guide the capillary assembly of soft- and hard colloids, and to link well-defined arrangements of polystyrene or silica particles produced with capillary assembly, a process we term "printing on particles". The printing process is based on automated particle recognition algorithms and enables the user to connect colloids into one- and two-dimensional tailored structures, including particle clusters and lattices of varying symmetry and composition, using commercial photo-resists (IP-L or IP-PDMS). Once printed and developed, the structures can be easily harvested and re-dispersed in water. The flexibility of our method allows the combination of a wide range of materials into complex structures, which we envisage will boost the realization of new systems for a broad range of fields, including microrobotics, micromanipulation and metamaterials.  
\end{abstract}


\section{Introduction}

3D printing has become a wide-spread technique to realize complex parts and intricate structures using a broad range of materials, from elastomers\cite{Guo2020}, metals\cite{Hirt2017} and glass\cite{Wen2021} to the direct printing of emulsions\cite{Minas2016}, cells\cite{Dey2020} and bacterial suspensions \cite{Schaffner2017}. 

When it comes to research applications, 3D printing has opened up many interesting avenues to provide new platforms for the realization of materials with high structural control. Furthermore, the ever-increasing demand for miniaturization and the desire to structure materials at the (sub)micron scale has driven the development of micro and nano-printing techniques. Among those, two-photon polymerization (2PP) 3D printing is a direct laser writing (DLW) technique that affords exquisite spatial resolution in the 100 nm range.\cite{HARINARAYANA2021107180} However, this miniaturization comes at the cost of a reduced choice of printable materials, typically limited to a handful of organic inks and photoresists.  \cite{Hippler2019,Nishiguchi2020,Kotz2021,Dayan2021,Babakhanova2018} In spite of the great progress, significant challenges remain. In particular, the integration and precise placement of multiple and distinct materials, e.g. organic and inorganic ones, in a single microscale printing process are currently elusive. Examples have shown the dispersion of nanoparticles within a photoresist or the post-modification of the printed structures by deposition and/or plating processes.\cite{Nishiguchi2020,Doherty2020,Zheng2020} However, these approaches do not enable micrometric spatial control on the localization of the different materials, for which there is only a limited range of possible choices. 
Nonetheless, the combination of inorganic and organic, hard and soft components, dynamic and static materials, would enable many new directions of research, e.g. into metamaterials.\cite{Qu2017} \\

The realization of complex 2D and 3D materials with precise microstructural control has conversely been at the core of large efforts in the field of particle synthesis and assembly. Colloidal synthesis routes offer a huge palette of particles of different materials, with exquisite control on shape and functionality. However, their self-assembly within large scale structures presents problems due to the necessity of controlling interactions in a very subtle and precise way, and success has been achieved only in few cases\cite{Wang2015,He2020}. Moreover, the microstructures obtained by self-assembly are mostly limited to close-packed ones following energy minimization.\cite{Gong2017,Kraft2009,Youssef2016} Therefore, composite colloidal structures fabricated via spontaneous assembly have a limited range of configurations, unless directionality is built into prescribed surface properties, i.e. via patchiness\cite{Duguet2016}. However, the realization and robust assembly of patchy particles with different interaction symmetries is highly challenging and typically only structures containing a few particles are obtained. 
To bypass these issues, the directed assembly of colloidal particles within (micro-)templates can be used.\cite{Yin2001,Ni2018,Malaquin2007} However, these approaches only enable the realization of structures via the packing of preformed particles, which, compared to the intricacy and design flexibility of directly printed structures, show clear limitations. 

In this manuscript, we propose an alternative strategy, which combines the best features of direct printing and of the assembly of pre-synthesized particles to produce multi-material structures with precise composition and versatile geometry.\\ 

\begin{figure}
  \centering
  \includegraphics[width=\linewidth]{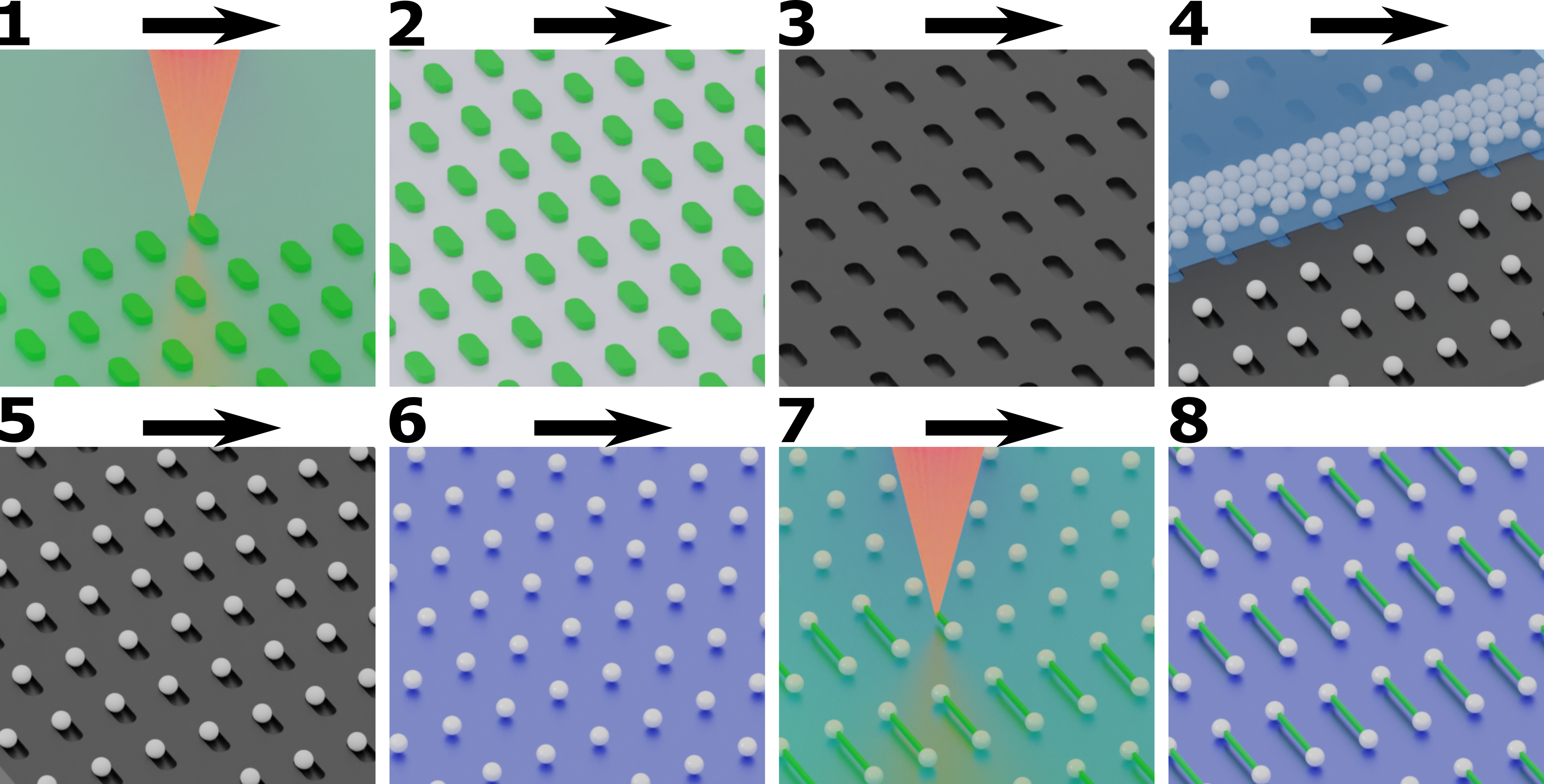}
  \caption{Schematic overview of the process flow for "printing-on-particles".  1. 2PP DLW printing of CAPA masters 2. Masters printed with 2PP DLW. 3. PDMS negative of the master to make the traps. 4. CAPA Deposition of PS or SiO\textsubscript{2} microparticles. 5. Traps filled with microparticles. 6. Transfer to a sacrificial layer. 7. Printing links between the microparticles with 2PP DWL. 8. Final linked structures that can be released by dissolving the sacrificial layer. }
  \label{fig:fig1}
\end{figure}

We achieve this goal via the union of 2PP 3D-printing and capillarity-assisted particle assembly (CAPA). In our process flow, the two approaches are closely intertwined (Figure \ref{fig:fig1}). We start by using DLW to fabricate masters of the templates used for CAPA. The  structuring of the height profiles of each individual cavity, or trap, used for the assembly enables the precise spatial positioning of particles of different sizes and mechanical properties. We then use CAPA to produce large-scale (mm\textsuperscript{2}) particle arrays with well-defined symmetry and placement of different particles, e.g. organic and inorganic ones. Finally, we use 2PP nanolithography again to print onto the deposited particles to modify their shape, or to link them into complex colloidal structures, from various '' colloidal molecules''\cite{Li2020} to larger lattices. We term this process "printing-on-particles" and show that it can be performed on multiple types of particles and using different printing materials. Finally, we present an easy harvesting procedure for the printed structures, which allows their redispersion and use.

\section{Results}

\subsection{3D-printed capillary assembly traps for multi-material colloids}

We begin by illustrating how 2PP can be used to expand the range of possible colloidal structures obtained by CAPA. CAPA is an assembly method where particles suspended in an evaporating droplet are selectively deposited inside microfabricated traps during the receding motion of the droplet's meniscus over an array of such traps, i.e. the template. In particular, the motion of the droplet's meniscus is typically driven by moving the substrate under a pinned droplet. Evaporation causes an accumulation of particles at the meniscus and capillary forces push the particles inside the traps as the droplet moves over them, leaving them inside the template upon depinning of the contact line\cite{Ni2015,Yin2001}. Conventional CAPA has been widely used as a patterning tool\cite{Malaquin2007}, but it was recently expanded into fabricating programmable multi-material colloidal structures via sequential deposition steps, or sCAPA\cite{Ni2016}. In sCAPA, a precise control of the trap depth profile is required to ensure that during each deposition step only one particle is deposited. The key to sCAPA is to tune the magnitude and direction of the capillary force relative to the particle position in the trap. In particular, the optimal trap depth is found to be around the radius of the deposited colloid\cite{Ni2015,Ni2016}. This poses a challenge if one wants to assemble colloids of different sizes or different stiffness, where soft particles can be deformed during deposition. Precise placement of different particles in different parts of the trap then requires a 3D height profile of each trap, with different heights for the different colloids. However, most common photolithography techniques used to fabricate the template generate features of a uniform depth and the fabrication of multiple heights in one trap would require multiple alignment, exposure and development steps. In contrast, 2PP is perfectly suited to print structures with non-uniform height profiles and holds the key to enable our multi-material colloidal structures.\\


\begin{figure}
  \centering
  \includegraphics[width=\linewidth]{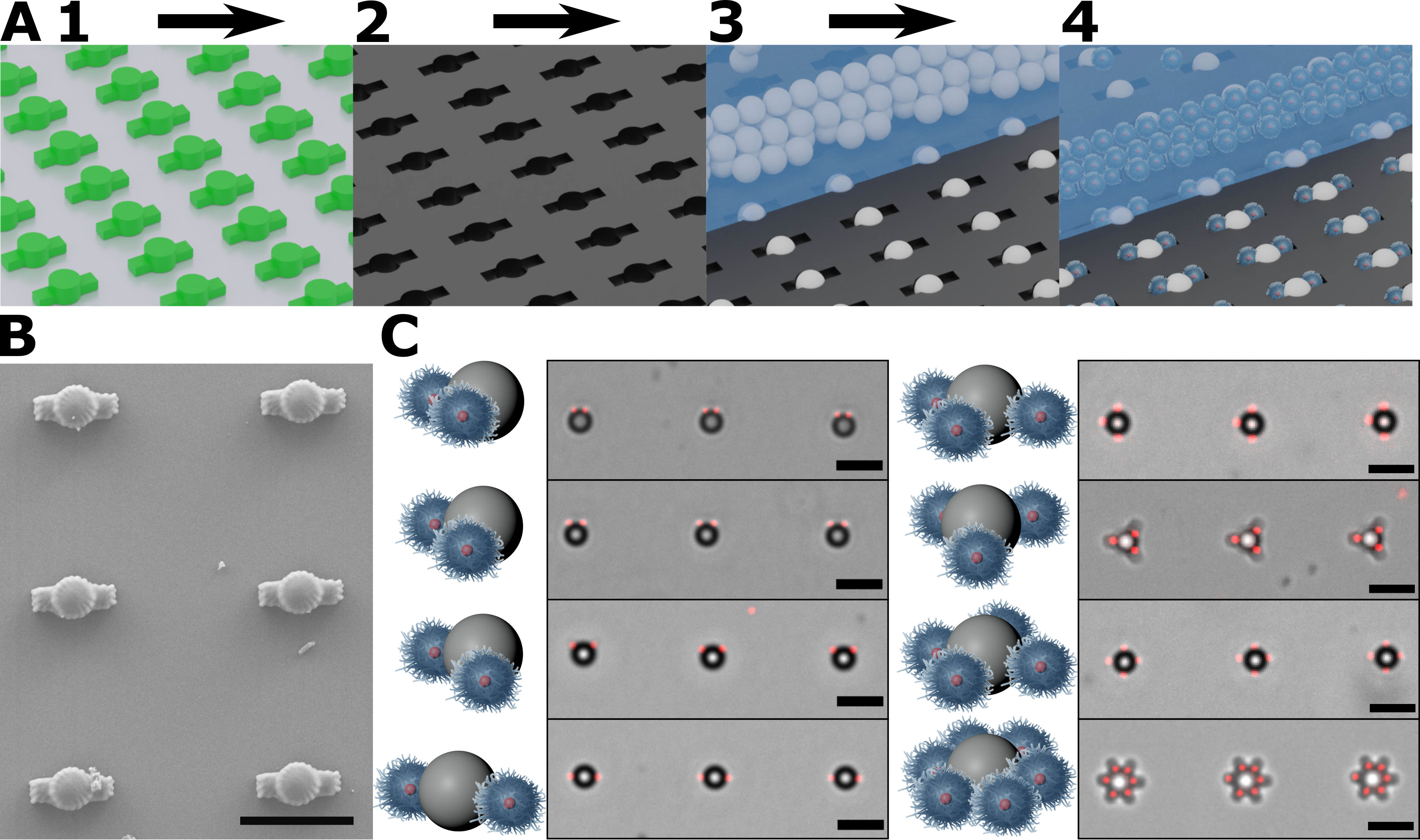}
  \caption{3D-printed traps for sCAPA of soft-hard colloidal clusters. (A) Schematic overview of the steps to fabricate soft-hard colloidal clusters. 1. Masters printed with 2PP DWL. 2. PDMS negative of the master to make the traps. 3. Deposition of hard 2 \textmu m PS particles. 4. Deposition of soft PNIPAM microgels. (B) SEM micrograph of traps printed with 2PP DWL for with 180{$^\circ$} three-particle clusters. (C) Bright-field + fluorescence micrographs of multiple clusters of 2 \textmu m PS and 2 \textmu m PNIPAM microgels with red fluorescent cores. Scale bars are 5 \textmu m.}
  \label{fig:fig2} 
\end{figure}

In particular, here we demonstrate the realization of hybrid colloidal molecules comprising soft responsive hydrogel particles (microgels) and hard polystyrene particles that could not be attained with 2.5D trap designs. Previous integration of microgels in sCAPA-fabricated colloidal clusters using traps with uniform depth profiles obtained by conventional lithography was limited to dumbbells and other simple shapes\cite{Alvarez2021}.
The range of structures that can be obtained with 2PP-printed traps is significantly expanded, as shown in Figure \ref{fig:fig2}.\\

Here, we use polystyrene particles with a 2 \textmu m diameter and poly(isopropyl-acrylamide) (PNIPAM)-based microgels with a red-fluorescence PTFMA core (see Methods for details). The PNIPAM-based microgels also have a nominal hydrodynamic diameter of 2 \textmu m, however they are deformed during the capillary assembly and can squeeze into traps of much smaller size. 


For the successful co-assembly of the two particle types, we have designed masters with a 1.5 \textmu m tall cylinder with 2.4 \textmu m diameter that fits the larger polystyrene colloids. Around this cylinder 1.2 \textmu m wide, 1.5 \textmu m long, 0.8 \textmu m tall, rectangular cuboid “arms” are placed that can fit the microgels (Figure \ref{fig:fig2}B) (CAD drawings can be found in Supplementary section 1.1). These masters are then replicated into PDMS to obtain the traps used in sCAPA. By changing number of arms and their relative positions, a host of different complex colloidal molecules are conceived and realized (Figure \ref{fig:fig2}C). These hybrid structures can be directly used to obtain responsive microswimmers. Trap designs with arms of different sizes allow for the co-assembly of multiple microgels which behave as microswimmers with multiple dynamical states.   \\

\subsection{Printed colloidal molecules}

The next step is to directly use 2PP-DLW in combination with sCAPA to print colloidal molecules. Opposed to printing whole colloidal-scale objects as monolithic structures composed only of the printed resin\cite{Doherty2020,Lee2021}, sCAPA enables the precise positioning of colloidal particles in prescribed arrangements and their subsequent joining by 3D-printing polymeric links to realize composite structures. 

The process to link colloids is briefly described here, and a more detailed description can be found in the Materials and Methods section and in the SI. We start with colloids presenting acrylate or methacrylate surface groups. The presence of these functional groups enables the chemical binding of the resin to the surface of the particles during the printing process, differently to the case of physical connections, e.g. where the particle is simply encased in a printed polymeric shell\cite{Hu2021,Xu2018}. Strong binding between the particles and the printed links is important to ensure minimal damage during the development and harvesting steps described below. We then deposit the colloids of interest into the PDMS templates by means of sCAPA, as schematically shown in Figure \ref{fig:fig1}-4 (trap sizes are shown in Supplementary section 1.2). This deposition method allows for the rapid patterning of arrays of colloids over cm\textsuperscript{2} areas, corresponding to $\approx 10^{6}$ particles, with deposition yields > 99\% (Figure \ref{fig:fig1}-5). The array of colloids is then transferred from the PDMS template to a slide for 2PP-DLW by means of an adhesive water-soluble sacrificial layer (glucose-dextran 1:1) (Figure \ref{fig:fig1}-6). Subsequently, the colloids are immersed in the photosensitive resin (IP-L, Nanoscribe Gmbh). Crucially, the sacrificial layer does not dissolve in the resin and keeps the transferred particles fixed during the printing step (Figure \ref{fig:fig1}-7). After the printing process is completed, the linked particles are developed in isopropyl alcohol (IPA) for 10 minutes and then dried in supercritical CO\textsubscript{2}. The sacrificial layer is unaffected by this process.   \\

\begin{figure}
  \centering
  \includegraphics[width=\linewidth]{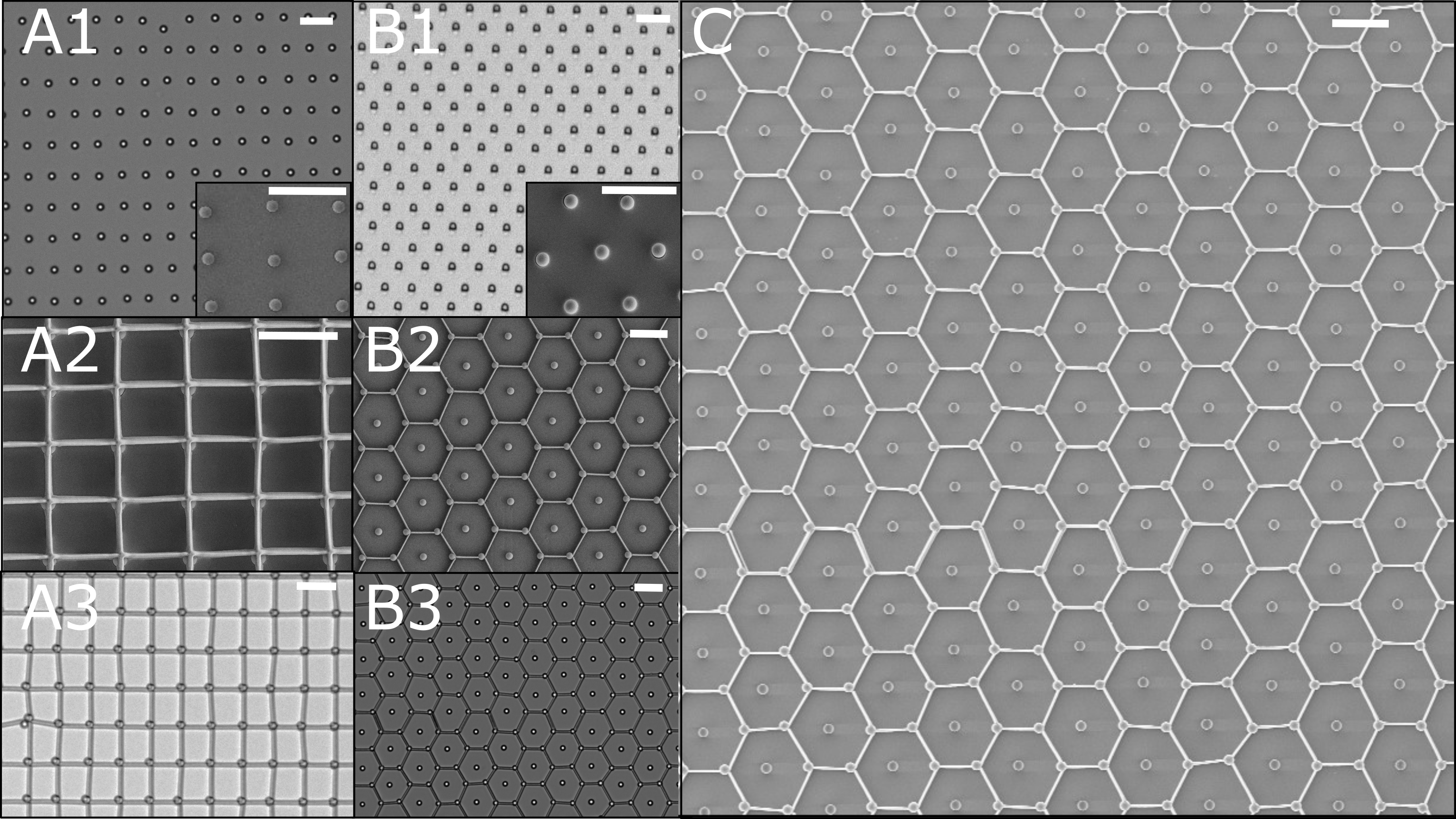}
  \caption{ Colloidal lattices fabricated with 2PP-DLW: A) Fabrication of rectangular colloidal lattices: A1-optical micrograph of 2 \textmu m PS particles deposited in a rectangular pattern by CAPA. Inset: a zoomed-in SEM micrograph of the same. A2-SEM and A3) optical microscopic images of rectangular lattices linked with IP-L after CPD. B) Fabrication of hexagonal colloidal lattices: B1-optical micrograph of 2 \textmu m SiO\textsubscript{2} particles deposited in a hexagonal pattern. Inset: a zoomed-in SEM micrograph of the same.  B2-SEM and B3 optical microscopic images of linked hexagonal colloidal lattices  with IP-L after CPD. (C) SEM micrograph of a hexagonal lattice over a larger view. Scale bars, 10 \textmu m.}
  \label{fig:fig4}
\end{figure}

The presence of regular particle arrays enables a fully automated printing process based on an algorithm written in Python for the detection and linking of the particles. The 2PP-DLW setup (Nanoscribe GT2) uses a conventional inverted microscope to project the NIR laser light required for the crosslinking of the resin and includes a camera to monitor the printing. We use this camera to take an image of a colloidal array on the substrate. Then, we employ a particle tracking package (TrackPy\cite{TrackPy,Crocker1995}) to locate each particle in the image and perform a nearest-neighbors search to find the immediate neighbors of each particle and label them based on their relative position. Subsequently, the user chooses which ones of these neighbors are to be linked and which particles are removed from the look-up table. Particles can be connected in a large variety of simple individual shapes ( e.g. see SI Figure 4-C1 for rectangular units or SI Figure-C2 for triangular ones) or joined into large-scale 2D lattices, as shown in Figure \ref{fig:fig4}. The set of links is converted to a print file that can be read by the DLW printer, such that the linking program dynamically creates files to control stage movement, image acquisition and printing. After an initial setup and calibration, the algorithm can be used to automatically connect particles deposited by sCAPA over large areas, up to a few mm\textsubscript{2}. \\

The after printing, the samples are developed in IPA and dried in supercritical CO\textsubscript{2} to eliminate capillary forces during drying that can break the links. Example colloidal molecules can be found in Supplementary section 4 and 5. Figure \ref{fig:fig4}A1-B2 show examples of rectangular and hexagonal lattices comprising 2 $\mu$m SiO\textsubscript{2} particles linked together with thin $\approx 500$ nm-thick lines of IP-L polymer resin after critical point drying (CPD). These structures are smaller than 10 \textmu m in size and have a characteristic "ball and stick molecule" appearance, suggesting analogies with so-called "colloidal molecules".\\

We then expand on this basic process by printing continuous lattices, of which we control the symmetry over large areas starting from particle arrays deposited via CAPA with a yield >99\% over squared millimeters. Figure \ref{fig:fig4}A) shows rectangular arrays of PS particles linked into rectangular lattices and Figure \ref{fig:fig4}B-C display  SiO\textsubscript{2} colloids deposited in a hexagonal pattern and linked into a honeycomb structure. 
During printing, only particles within a field of view of approximately 200 \textmu m x 140 \textmu m  can be linked and the continuous lattices are obtained by stitching many fields of view together.


\subsection{Three-material structures}
\begin{figure}
  \centering
  \includegraphics[width=\linewidth]{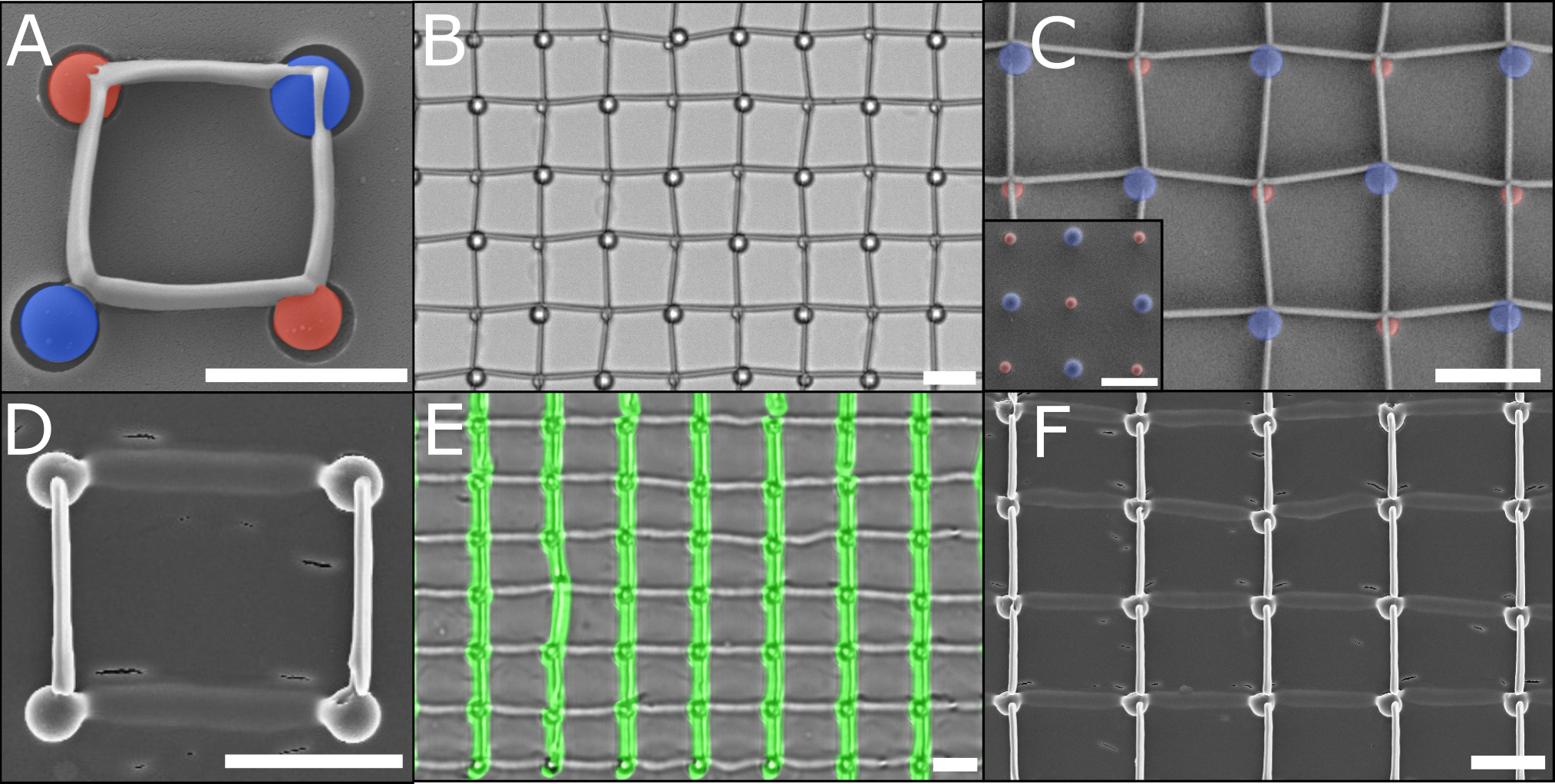}
  \caption{Three-material structures: (A) SEM image of a rectangular "colloidal molecule" obtained from 2 \textmu m PS and SiO\textsubscript{2} particles linked with IP-L. (B-C) Bright-field microscopy image (B) and SEM image (C) of three-material colloidal lattices fabricated from 3 \textmu m PS and 2 \textmu m SiO\textsubscript{2} colloids linked with IP-L into a 2D rock-salt structure. (D) SEM image of a rectangular three-material "colloidal molecule" fabricated linking SiO\textsubscript{2} particles with IP-L (vertical) and IP-PDMS (horizontal) resins. (E) Composite bright-field and fluorescence microscopy image and SEM image (F) of a rectangular three-material colloidal lattice fabricated linking SiO\textsubscript{2} particles with IP-L (vertical) and IP-PDMS (horizontal) resins, Note that the PS particles are false-colored in blue and the SiO\textsubscript{2} particles in red to visually distinguish the two materials in SEM images (A and C). Scale bars, 5 \textmu m.}
  \label{fig:fig5}
\end{figure}

The core advantage of combining sCAPA with 2PP-DLW is the precise incorporation of different materials into fabricated structures, which can be extended beyond the two-materials examples reported above. In particular, sCAPA allows for the deposition colloids of different materials in prescribed sequences and positions. Figure \ref{fig:fig5}A-C show the results of creating and linking mixed arrays of PS and SiO\textsubscript{2} colloids. The two types of particles are first deposited in traps of matching sizes in two deposition steps (the larger PS particles, false-colored in blue are deposited first), transferred as previously described and finally linked using IP-L into either multi-material small "molecules" (Figure \ref{fig:fig5}A) or into large arrays with a 2D rock salt arrangement (Figure \ref{fig:fig5}B-C). 

Complementary, multiple materials can instead be incorporated by applying, printing and developing different resins in a sequence.\cite{Puce2019,Ceylan2017} Here, we demonstrate that the same concept can be extend to linking colloids. Crucially, the particle adhesion layer must be resistant to the developing steps of all resins in order to preserve the structures. In Figure \ref{fig:fig4} we demonstrate rectangular "colloidal molecules" linking SiO\textsubscript{2} colloids. We first printed horizontal links with IP-PDMS, a silicone elastomer-based resin, and then developed the sample. We subsequently re-immersed the sample in IP-L to print the vertical links obtaining structures where hard and soft links are combined (Figure \ref{fig:fig5}D). As previously shown, these small-scale structures can be expanded into large lattices by connecting particles with continuous lines and stitching multiple fields of view together (Figure \ref{fig:fig5}E-F). 

\subsection{Shaping links and harvesting colloidal molecules}

\begin{figure}
  \centering
  \includegraphics[width=\linewidth]{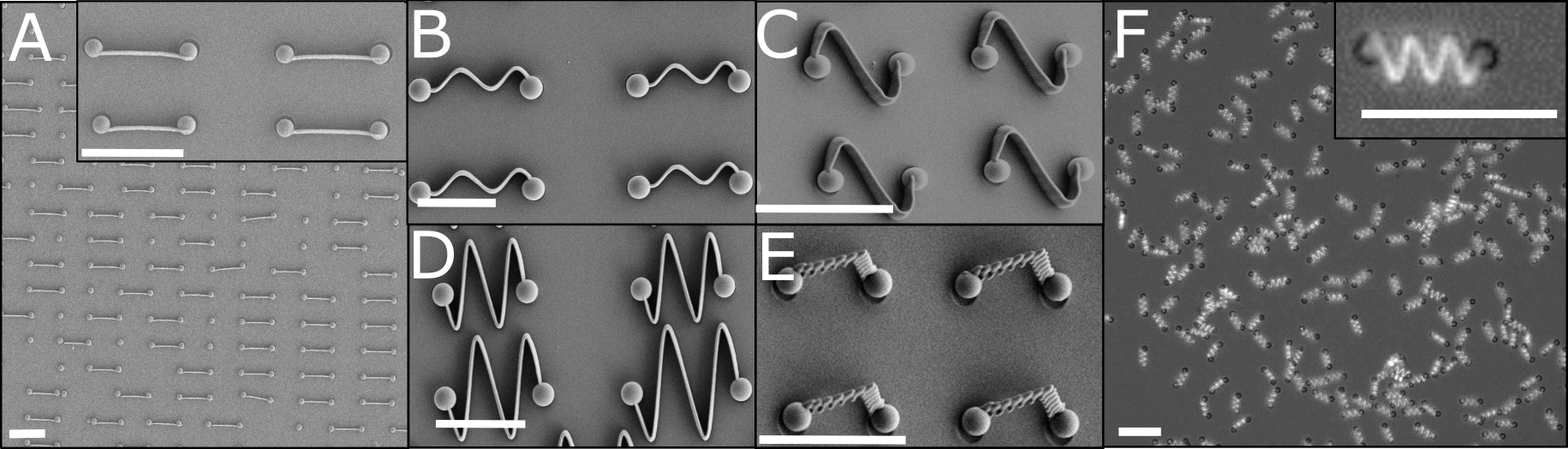}
  \caption{Fabrication and harvesting of colloidal molecules: (A-E) SEM images of colloidal PS dimers linked with IP-L in various shapes, from straight lines (A), to sinusoidal links of varying amplitude and frequency attached to the particles with a 45 degrees angle (B-D). The thickness of the printed links is 0.5 \textmu m. (E) Tilted SEM image of out-of-plane links connecting neighboring 2 \textmu m PS particles. The discontinuities in the print correspond to the smallest polymerizable volume (voxel) in the 2PP-DLW process. The highest point of the links is 1 \textmu m higher in the z-direction relative to the center of PS particles. F) Combined  bright-field and fluorescence microscopy image of harvested colloidal dimers consisting of two 2 \textmu m PS particles linked with sine-waved IP-L after release in water. Inset: Magnified colloidal dimer. Scale bars, 10 \textmu m.}
  \label{fig:fig6}
\end{figure}

Finally, we show that different shapes of the links can be realized beyond the straight ones showed so far. In fact, our automated Python printing code enables full control over the laser path to make more complex shapes at high resolution. As examples, we used sine waves with different wavelength and amplitudes to define the printing path (Figure \ref{fig:fig6}B,D,E). Furthermore, by exploiting the exquisite 3D resolution of 2PP-DLW in three dimensions, we have also demonstrated the printing of out-of-plane links between the deposited colloids (Figure \ref{fig:fig6}C). 

The deliberate choice of a dextran-glucose sacrificial layer, ultimately makes the release of the printed structures a very easy step.  By simply adding an aqueous solution to the substrate dissolves the adhesion layer and releases the structures into aqueous environments for further uses. Figure \ref{fig:fig6}F shows a large, homogeneous population of colloidal dimers with sinusoidal links fully released and undergoing Brownian motion in an suspension (Figure \ref{fig:fig6}F, Supplementary Movie 1). The same harvesting process can also be extended to larger, millimeter-scale lattices (see Supplementary Section 6). As the IP-L links are hydrophobic, a small amount of the surfactant Triton-X45 is added to ensure colloidal stability of the harvested structures. \\

\section{Conclusions and Discussion}

In this manuscript we have demonstrated that the combination of 2PP-DLW and capillarity-assisted particle assembly is a powerful and versatile option to expand the palette of multi-material microstructures obtained via additive manufacturing. In particular, 2PP-DLW plays two roles in this process: first, it enables the realization of traps with three-dimensional profiles that hold the key for the co-assembly of microparticles with different dimensions and mechanical properties during sCAPA, and then is used for the direct printing of links that connect particles in specific 1D or 2D structures, from single units to millimeter-scale lattices. 

The level of control of 2PP-DLW comes nonethelss with a compromise on the scalability of the printing-on-particles process. The 2PP-DLW used in our study (Nanoscribe Photonic GT2) manufactures masters for sCAPA traps at a rate of 10\textsuperscript{4} traps per hour and it therefore is significantly slower than more traditional maskless photolithography processes, which can produce up to 10\textsuperscript{6} h\textsuperscript{-1}. The current printing speed is still sufficient to prepare a full sCAPA master within 16 h in an overnight print, but scaling up to greater production volumes remains challenging. Similarly, the printing-on-particles process is automated via a custom-written Python software and it allows printing links at a rate of 10\textsuperscript{5} h\textsuperscript{-1} with high reproducibility. However, the identification of the tracking parameters and the printing calibration have to be optimized for each different particle/resin system. In the future, the printing speed may be accelerated exploiting recent developments in holographic techniques that split the laser beam to allow parallel printing\cite{Maibohm2020} and a fully automated tracking and printing protocol is desirable. Therefore, we see a clear analogy with macro-scale additive manufacturing, which is best suited for rapid-prototyping and small-scale production.



In the future, we envisage extending our route to including functional materials in the 3D-printed links, such as liquid crystal elastomers or responsive hydrogels \cite{Guo2020,Hippler2019,Nishiguchi2020,Zheng2020,Pozo2022,He2019}, to be combined with different types of functional colloidal particles. Furthermore, slight alterations to the resin formulation may be considered, e.g. by adding charged monomers or steric stabilizers, like poly(ethylene glycol)-methacrylate (PEG-MA), to increase the colloidal stability of the harvested structures in water without the use of surfactants.



Finally, we expect a broad use of the printing-on-particles process to fabricate tailored asymmetric colloidal molecules for small-scale assembly and to realize complex functional microswimmers with a high degree of control over composition and the trajectories. \cite{Shields2017,Lee2021,Alvarez2021,Ceylan2017,Zeng2015,Elgeti2015} Intriguingly, the larger 2D colloidal lattices present similarities with 2D atomic materials such as graphene
or MoS\textsubscript{2}. We thus envisage that they can provide interesting analogies to study phononic transport and mechanical behavior at a larger scale, while providing new opportunities to design 2D colloidal metamaterials. \cite{Holmes2019} In conclusion, we believe that combining CAPA and 2PP-DLW for printing-on-particles opens up exciting possibilities for future reserach and adds a new tool in the palette of available small-scale additive manufacturing processes. 


\section{Materials and Methods}

All chemicals, if not stated otherwise, were used as provided by the supplier. Acrylic acid (AA), N, N' Methylenebis(acrylamide) (BIS), 2,2,2-Trifluoroethyl methacrylate (TFMA), Sodium dodecyl sulfate (SDS), Potassium persulfate (KPS) and Nile red were purchased from Sigma-Aldrich. N-Isopropylacrylamide (NIPAM) was purchased from Sigma-Aldrich and purified by recrystallization in Toluene/Hexane 50:50.

\subsection{Microgel synthesis}

The microgels with fluorescent cores and thermo-responsive shells were prepared in a two-step synthesis \cite{Go2014}:  first the PTFMA-cores containing a fluorescent dye were prepared using free radical emulsion polymerization, then a shell of PNIPAM and PAA co-polymer was grown by free radical precipitation polymerization around this core. For the cores, TFMA (10 mL; 12 g; 70,3 mmol), NIPAM (940 mg; 8,31 mmol), SDS (30 mg; 0,104 mmol) and Nile red (5 mg; 0.016 mmol) were added in milli-Q water (30 mL) and stirred at 600 rpm for 20 min and purged with N\textsubscript{2}. The reaction mixture was heated up to 70 \degree C. KPS (25 mg; 0.093 mmol), previously dissolved in water (2,5 mL), was added after the emulsion was stirred for 15 min at 70 \degree C. After 4 h of stirring at 70 \degree C the reaction was stopped by contact with air. The resulting particles were directly filtered. The purification of the particles was carried out via centrifugation followed by decantation, addition of Milli-Q water and particle redispersion. The purification via centrifugation was performed three times. To remove SDS residues, the redispersed particles were dialysed in water for 24 - 72 h (dialysis tube membrane: 12 - 14 kDa). The particles were analysed with dynamic light scattering (ZetaSizer Nano DLS).

The PNIPAM-co-AA shells were grown by first dissolving NIPAM(1.15g), BIS(11.5 mg) and AA(88.5 \textmu l) in 50 ml milli-Q water. Then 300 \textmu l of the core particle suspension was added. The mixture was stirred at 600 rpm for 20 min and purged with N\textsubscript{2}. The reaction mixture was heated up to 70 \degree C and KPS (25 mg; 0.093 mmol) was slowly added. After 2 h, the reaction reached completion and the microgel suspension was filtered and dialysed for 48h  (dialysis tube membrane: 12 - 14 kDa). The particles were characterized with DLS.

\subsection{sCAPA trap design}

Traps of different sizes and shapes were designed with a CAD program (Rhinoceros 3D). For CAPA depositions of 2 \textmu m SiO\textsubscript{2} and 2 \textmu m PS particles in rectanguar pattern, the traps are 2.2 \textmu m in width, 4.2 \textmu m in length and 1 \textmu m in height. CAPA traps designed for the deposition of 2 \textmu m SiO\textsubscript{2} as well as 2 \textmu m PS particles in hexagonal patterns have the dimensions: 1.25 \textmu m in length, 1 \textmu m in height, and 2.5 \textmu m in width. Longer sCAPA traps were designed for the double deposition of 2 \textmu m PS and 2 \textmu m SiO\textsubscript{2} in opposite directions. These traps are composed of a cuboid and semi-cylinders on both ends of the cuboid. The dimensions of the trap are: 0.7 \textmu m in height, 10 \textmu m in length, and width of 1.2 \textmu m. Double-deposition sCAPA traps for 3 \textmu m PS and 2 \textmu m SiO\textsubscript{2} were designed by combining single-particle traps for 3 \textmu m PS and 2 \textmu m SiO\textsubscript{2} in an array. The single-particle traps for 3 \textmu m PS have the same shape as the 2 \textmu m SiO\textsubscript{2} particles, with dimension of: 2 \textmu m in length, 1.5 \textmu m in height and 4 \textmu m in width. The schemes for each type of trap with dimensions are found in the SI Figure 2

\subsection{Printing sCAPA traps with NanoScribe}

A fused silica substrate (Multi-Dill, NanoScribe GmbH) was cleaned with the standard procedure from NanoScribe (rinsed with EtOH, plasma-treated for $\approx$ 30s).  A commercial Direct Laser Writing setup (Photonic Professional GT2, NanoScribe GmbH)  with a 63x, NA = 1.5 with commercial Dip-in resin (IP-Dip, NanoScribe) was used for template fabrication. The print is developed with a standard procedure (20 min in PGMEA, 10 min in IPA, rinsed with IPA) and post-cured under UV light (306 nm) for at least one hour. To decrease the adhesion of glass substrate and the microstructures, Trichloro(1H,1H,2H,2H-perfluorooctyl) silane (Sigma-Aldrich, \textgreater 97 \%) is coated on the substrate by means of chemical vapor deposition, by exposing the substrate to a 100 \textmu l droplet of the silane for 30 min and rinsed with ethanol afterward. PDMS (SYLGARD 184 silicone elastomer, Dow Chemicals) was poured over the silica substrate and cured overnight at 65 $^\circ C$. Finally, the cured PDMS was removed from the master and used as a template for sCAPA deposition.

\subsection{Particle functionalization and sCAPA deposition}

The 2 \textmu m silica (SiO\textsubscript{2}) particles (Microparticles GmbH) were functionalized with methacrylate groups to enhance adhesion between colloids and the printed structure. The functionalization was performed as follows. First, 500 \textmu l of 2 \textmu m aqueous SiO\textsubscript{2} particle solution at 5 w/w\% were repeatedly centrifuged and redispersed in MilliQ water. Afterwards, 500 \textmu L of hydrogen peroxide (Sigma-Aldrich, 30 \%) and 500 \textmu l of ammonia solution (25 \% w/w) were mixed with the particle suspension for 10 min in an oil bath at 70 $^\circ C$. This suspension was again washed with MilliQ water and brought to a volume of 1 mL. The solution was stirred overnight with 4 mL of dry ethanol (Sigma-Aldrich, >99.8 \%) and 25 \textmu l of 3-(trimethoxysilyl)propyl methacrylate (Sigma-Aldrich, >98 \%) and finally washed again by centrifugation and redispersion with Milli-Q water. The 2 and 3 \textmu m polystyrene particles were already functionalized upon purchase (Micromod Partikeltechnologie GmbH). Before deposition, the functionalized particles were dispersed in a deposition solution composed of Triton X-45 (Sigma-Aldrich, \textgreater 99.8 \%), SDS (\textgreater 99.8 \%), Milli Q water, (0.08 M) NaOH, and glycerol, to a concentration specified in SI. Suspensions of different particles were freshly prepared prior to each deposition and sonicated for 10 minutes to avoid agglomeration. Around 60 \textmu l of deposition suspension was added over the PDMS template and dragged along at a constant speed by a flat PDMS slab at a speed of 3-5\textmu m s\textsuperscript{-1} and at a temperature of 25 $^\circ C$. sCAPA in long traps was done by depositing suspensions of 2 \textmu m SiO\textsubscript{2} particles in one direction and 2 \textmu m PS particles in the opposite direction. Deposition of mixed arrays of 3 \textmu m PS and 2 \textmu m SiO\textsubscript{2} was done by depositing the 3 \textmu m PS particles and subsequently the 2 \textmu m SiO\textsubscript{2} in the same direction.

\subsection{Fabrication and harvesting of micro-structures}

A borosilicate glass substrate (30 mm, Thermo Scientific) was rinsed with ethanol and plasma treated. A solution of dextran (20 w/v \%) and glucose (20 w/v \%) was dissolved in Milli Q water and spin-coated at 4000 rpm for 15 s. After spin coating, the PDMS template with the deposited particles was pressed and gently peeled off to transfer the particles onto the glass substrate. A drop of IPL resin was placed on the substrate and covered the region of the transferred particles. After printing, the prints were washed with IPA and critical point dried (EM CPD3000, Leica Microsystems). For harvesting, a microscope spacer was placed around the critical point dried prints and closed off with a smaller glass slide (10 mm, Thermo Scientific) forming a cavity. Successively, 10 \textmu l Milli Q water was pumped into the cavity to dissolve the dextran-glucose layer and release the printed structures. The harvesting procedure was recorded with an optical microscope (Eclipse Ti-2, Nikon).

\subsection{Imaging}

The particles linked by commercial photo-resist (IPL, NanoScribe GmbH) were imaged by bright-field and fluorescence optical microscopy (Nikon, Eclipse Ti-2) at various magnifications and by SEM (SU8000, Hitachi). 

\section*{Acknowledgments}
We thank Minghan Hu, Laura Alvarez and Simon Scherrer for the SEM images. This project has received funding from the European Research Council (ERC) under the the European Union’s Horizon 2020 research and innovation program grant agreement No 101001514.

\bibliographystyle{unsrt}  
\bibliography{references}  

\begin{thebibliography}{10}

\bibitem{Guo2020}
Yubing Guo, Hamed Shahsavan, and Metin Sitti.
\newblock 3d microstructures of liquid crystal networks with programmed
  voxelated director fields.
\newblock {\em Advanced Materials}, 32:1--10, 2020.

\bibitem{Hirt2017}
Luca Hirt, Alain Reiser, Ralph Spolenak, and Tomaso Zambelli.
\newblock Additive manufacturing of metal structures at the micrometer scale.
\newblock {\em Advanced Materials}, 29:1604211, 5 2017.

\bibitem{Wen2021}
Xiewen Wen, Boyu Zhang, Weipeng Wang, Fan Ye, Shuai Yue, Hua Guo, Guanhui Gao,
  Yushun Zhao, Qiyi Fang, Christine Nguyen, Xiang Zhang, Jiming Bao, Jacob~T.
  Robinson, Pulickel~M. Ajayan, and Jun Lou.
\newblock 3d-printed silica with nanoscale resolution.
\newblock {\em Nature Materials 2021 20:11}, 20:1506--1511, 10 2021.

\bibitem{Minas2016}
Clara Minas, Davide Carnelli, Elena Tervoort, André~R Studart, C~Minas,
  D~Carnelli, E~Tervoort, and A~R Studart.
\newblock 3d printing of emulsions and foams into hierarchical porous ceramics.
\newblock {\em Advanced Materials}, 28:9993--9999, 12 2016.

\bibitem{Dey2020}
Madhuri Dey and Ibrahim~T. Ozbolat.
\newblock 3d bioprinting of cells, tissues and organs.
\newblock {\em Scientific Reports 2020 10:1}, 10:1--3, 8 2020.

\bibitem{Schaffner2017}
Manuel Schaffner, Patrick~A. R{\"{u}}hs, Fergal Coulter, Samuel Kilcher, and
  Andr{\'{e}}~R. Studart.
\newblock {3D printing of bacteria into functional complex materials}.
\newblock {\em Science Advances}, 3(12), 2017.

\bibitem{HARINARAYANA2021107180}
V.~Harinarayana and Y.C. Shin.
\newblock Two-photon lithography for three-dimensional fabrication in
  micro/nanoscale regime: A comprehensive review.
\newblock {\em Optics \& Laser Technology}, 142:107180, 2021.

\bibitem{Hippler2019}
Marc Hippler, Eva Blasco, Jingyuan Qu, Motomu Tanaka, Christopher
  Barner-Kowollik, Martin Wegener, and Martin Bastmeyer.
\newblock Controlling the shape of 3d microstructures by temperature and light.
\newblock {\em Nature Communications}, 10, 12 2019.

\bibitem{Nishiguchi2020}
Akihiro Nishiguchi, Hang Zhang, Sjören Schweizerhof, Marie~Friederike Schulte,
  Ahmed Mourran, and Martin Möller.
\newblock 4d printing of a light-driven soft actuator with programmed printing
  density.
\newblock {\em ACS Applied Materials and Interfaces}, 12:12176--12185, 2020.

\bibitem{Kotz2021}
Frederik Kotz, Alexander~S. Quick, Patrick Risch, Tanja Martin, Tobias Hoose,
  Michael Thiel, Dorothea Helmer, and Bastian~E. Rapp.
\newblock {Two-Photon Polymerization of Nanocomposites for the Fabrication of
  Transparent Fused Silica Glass Microstructures}.
\newblock {\em Advanced Materials}, 33(9), 2021.

\bibitem{Dayan2021}
Cem~Balda Dayan, Sungwoo Chun, Nagaraj Krishna-Subbaiah, Dirk~Michael Drotlef,
  Mukrime~Birgul Akolpoglu, and Metin Sitti.
\newblock {3D Printing of Elastomeric Bioinspired Complex Adhesive
  Microstructures}.
\newblock {\em Advanced Materials}, 33(40), oct 2021.

\bibitem{Babakhanova2018}
Greta Babakhanova, Taras Turiv, Yubing Guo, Matthew Hendrikx, Qi-Huo Wei,
  Albert P. H.~J. Schenning, Dirk~J. Broer, and Oleg~D. Lavrentovich.
\newblock Liquid crystal elastomer coatings with programmed response of surface
  profile.
\newblock {\em Nature Communications}, 9(1):456, Jan 2018.

\bibitem{Doherty2020}
Rachel~P. Doherty, Thijs Varkevisser, Margot Teunisse, Jonas Hoecht, Stefania
  Ketzetzi, Samia Ouhajji, and Daniela~J. Kraft.
\newblock Catalytically propelled 3d printed colloidal microswimmers.
\newblock {\em Soft Matter}, 16:10463--10469, 12 2020.

\bibitem{Zheng2020}
Chenglin Zheng, Feng Jin, Yuanyuan Zhao, Meiling Zheng, Jie Liu, Xianzi Dong,
  Zhong Xiong, Yanzhi Xia, and Xuanming Duan.
\newblock {Light-driven micron-scale 3D hydrogel actuator produced by
  two-photon polymerization microfabrication}.
\newblock {\em Sensors and Actuators B: Chemical}, 304:127345, feb 2020.

\bibitem{Qu2017}
Jingyuan Qu, Muamer Kadic, Andreas Naber, and Martin Wegener.
\newblock Micro-structured two-component 3d metamaterials with negative
  thermal-expansion coefficient from positive constituents.
\newblock {\em Scientific Reports}, 7, 1 2017.

\bibitem{Wang2015}
Yu~Wang, Yufeng Wang, Xiaolong Zheng, Étienne Ducrot, Jeremy~S. Yodh, Marcus
  Weck, and David~J. Pine.
\newblock Crystallization of dna-coated colloids.
\newblock {\em Nature Communications 2015 6:1}, 6:1--8, 6 2015.

\bibitem{He2020}
Mingxin He, Johnathon~P. Gales, Étienne Ducrot, Zhe Gong, Gi~Ra Yi, Stefano
  Sacanna, and David~J. Pine.
\newblock Colloidal diamond.
\newblock {\em Nature 2020 585:7826}, 585:524--529, 9 2020.

\bibitem{Gong2017}
Zhe Gong, Theodore Hueckel, Gi~Ra Yi, and Stefano Sacanna.
\newblock Patchy particles made by colloidal fusion.
\newblock {\em Nature}, 550:234--238, 2017.

\bibitem{Kraft2009}
Daniela~J. Kraft, Jan Groenewold, and Willem~K. Kegel.
\newblock Colloidal molecules with well-controlled bond angles.
\newblock {\em Soft Matter}, 5:3823--3826, 2009.

\bibitem{Youssef2016}
Mena Youssef, Theodore Hueckel, Gi~Ra Yi, and Stefano Sacanna.
\newblock Shape-shifting colloids via stimulated dewetting.
\newblock {\em Nature Communications}, 7:1--7, 2016.

\bibitem{Duguet2016}
Étienne Duguet, Céline Hubert, Cyril Chomette, Adeline Perro, and Serge
  Ravaine.
\newblock Patchy colloidal particles for programmed self-assembly.
\newblock {\em Comptes Rendus Chimie}, 19:173--182, 1 2016.

\bibitem{Yin2001}
Y.~Yin, Y.~Lu, B.~Gates, and Y.~Xia.
\newblock Template-assisted self-assembly: A practical route to complex
  aggregates of monodispersed colloids with well-defined sizes, shapes, and
  structures.
\newblock {\em Journal of the American Chemical Society}, 123:8718--8729, 9
  2001.

\bibitem{Ni2018}
Songbo Ni, Lucio Isa, and Heiko Wolf.
\newblock Capillary assembly as a tool for the heterogeneous integration of
  micro- and nanoscale objects.
\newblock {\em Soft Matter}, 14:2978--2995, 2018.

\bibitem{Malaquin2007}
Laurent Malaquin, Tobias Kraus, Heinz Schmid, Emmanuel Delamarche, and Heiko
  Wolf.
\newblock Controlled particle placement through convective and capillary
  assembly.
\newblock {\em Langmuir}, 23:11513--11521, 11 2007.

\bibitem{Li2020}
Weiya Li, Hervé Palis, Rémi Mérindol, Jérôme Majimel, Serge Ravaine, and
  Etienne Duguet.
\newblock Colloidal molecules and patchy particles: complementary concepts,
  synthesis and self-assembly.
\newblock {\em Chemical Society Reviews}, 49:1955--1976, 3 2020.

\bibitem{Ni2015}
Songbo Ni, Jessica Leemann, Heiko Wolf, and Lucio Isa.
\newblock Insights into mechanisms of capillary assembly.
\newblock {\em Faraday Discussions}, 181:225--242, 2015.

\bibitem{Ni2016}
S.~Ni, J.~Leemann, I.~Buttinoni, L.~Isa, and H.~Wolf.
\newblock Programmable colloidal molecules from sequential capillarity-assisted
  particle assembly.
\newblock {\em Science Advances}, 2:e1501779--e1501779, 2016.

\bibitem{Alvarez2021}
L.~Alvarez, M.~A. Fernandez-Rodriguez, A.~Alegria, S.~Arrese-Igor, K.~Zhao,
  M.~Kröger, and Lucio Isa.
\newblock Reconfigurable artificial microswimmers with internal feedback.
\newblock {\em Nature Communications}, 12, 12 2021.

\bibitem{Lee2021}
Yun~Woo Lee, Hakan Ceylan, Immihan~Ceren Yasa, Ugur Kilic, and Metin Sitti.
\newblock 3d-printed multi-stimuli-responsive mobile micromachines.
\newblock {\em ACS Applied Materials and Interfaces}, 13:12759--12766, 2021.

\bibitem{Hu2021}
Xinghao Hu, Immihan~C. Yasa, Ziyu Ren, Sandhya~R. Goudu, Hakan Ceylan, Wenqi
  Hu, and Metin Sitti.
\newblock Magnetic soft micromachines made of linked microactuator networks.
\newblock {\em Science Advances}, 7, 6 2021.

\bibitem{Xu2018}
Bing Xu, Yang Shi, Zhaoxin Lao, Jincheng Ni, Guoqiang Li, Yanlei Hu, Jiawen Li,
  Jiaru Chu, Dong Wu, and Koji Sugioka.
\newblock Real-time two-photon lithography in controlled flow to create a
  single-microparticle array and particle-cluster array for optofluidic
  imaging.
\newblock {\em Lab on a Chip}, 18:442--450, 2 2018.

\bibitem{TrackPy}
Daniel~B. Allan, Thomas Caswell, Nathan~C. Keim, Casper~M. van~der Wel, and
  Ruben~W. Verweij.
\newblock soft-matter/trackpy: Trackpy v0.5.0, 2021.

\bibitem{Crocker1995}
John Crocker and David Grier.
\newblock Methods of digital video microscopy for colloidal studies | elsevier
  enhanced reader.
\newblock {\em Journal of Colloid and Interface Science}, 179:298--310, 1995.

\bibitem{Puce2019}
Salvatore Puce, Elisa Sciurti, Francesco Rizzi, Barbara Spagnolo, Antonio
  Qualtieri, Massimo {De Vittorio}, and Urs Staufer.
\newblock {3D-microfabrication by two-photon polymerization of an integrated
  sacrificial stencil mask}.
\newblock {\em Micro and Nano Engineering}, 2(January):70--75, 2019.

\bibitem{Ceylan2017}
Hakan Ceylan, Immihan~Ceren Yasa, and Metin Sitti.
\newblock {3D Chemical Patterning of Micromaterials for Encoded Functionality}.
\newblock {\em Advanced Materials}, 29(9):1605072, mar 2017.

\bibitem{Maibohm2020}
Christian Maibohm, Oscar~F. Silvestre, Jérôme Borme, Maina Sinou, Kevin
  Heggarty, and Jana~B. Nieder.
\newblock Multi-beam two-photon polymerization for fast large area 3d periodic
  structure fabrication for bioapplications.
\newblock {\em Scientific Reports 2020 10:1}, 10:1--10, 5 2020.

\bibitem{Pozo2022}
Marc del Pozo, Colm Delaney, Marina~Pilz da~Cunha, Michael~G. Debije, Larisa
  Florea, and Albert P. H.~J. Schenning.
\newblock Temperature‐responsive 4d liquid crystal microactuators fabricated
  by direct laser writing by two‐photon polymerization.
\newblock {\em Small Structures}, 3:2100158, 2 2022.

\bibitem{He2019}
Ziqian He, Guanjun Tan, Debashis Chanda, and Shin-Tson Wu.
\newblock Novel liquid crystal photonic devices enabled by two-photon
  polymerization [invited].
\newblock {\em Optics Express}, 27:11472, 4 2019.

\bibitem{Shields2017}
C.~Wyatt Shields and Orlin~D. Velev.
\newblock {The Evolution of Active Particles: Toward Externally Powered
  Self-Propelling and Self-Reconfiguring Particle Systems}.
\newblock {\em Chem}, 3(4):539--559, 2017.

\bibitem{Zeng2015}
Hao Zeng, Piotr Wasylczyk, Camilla Parmeggiani, Daniele Martella, Matteo
  Burresi, and Diederik~Sybolt Wiersma.
\newblock Light-fueled microscopic walkers.
\newblock {\em Advanced Materials}, 27:3883--3887, 7 2015.

\bibitem{Elgeti2015}
J.~Elgeti, R.~G. Winkler, and G.~Gompper.
\newblock Physics of microswimmers - single particle motion and collective
  behavior: A review.
\newblock {\em Reports on Progress in Physics}, 78, 2015.

\bibitem{Holmes2019}
Douglas~P. Holmes.
\newblock Elasticity and stability of shape-shifting structures.
\newblock {\em Current Opinion in Colloid and Interface Science}, 40:118--137,
  2019.

\end{thebibliography}

\end{document}


\maketitle{}

\tableofcontents

\newpage

\section{Trap designs}
\subsection{3D-traps for PNIPAM microgels}
\begin{figure}[h]
  
   \centering
  
  \includegraphics[width=\linewidth]{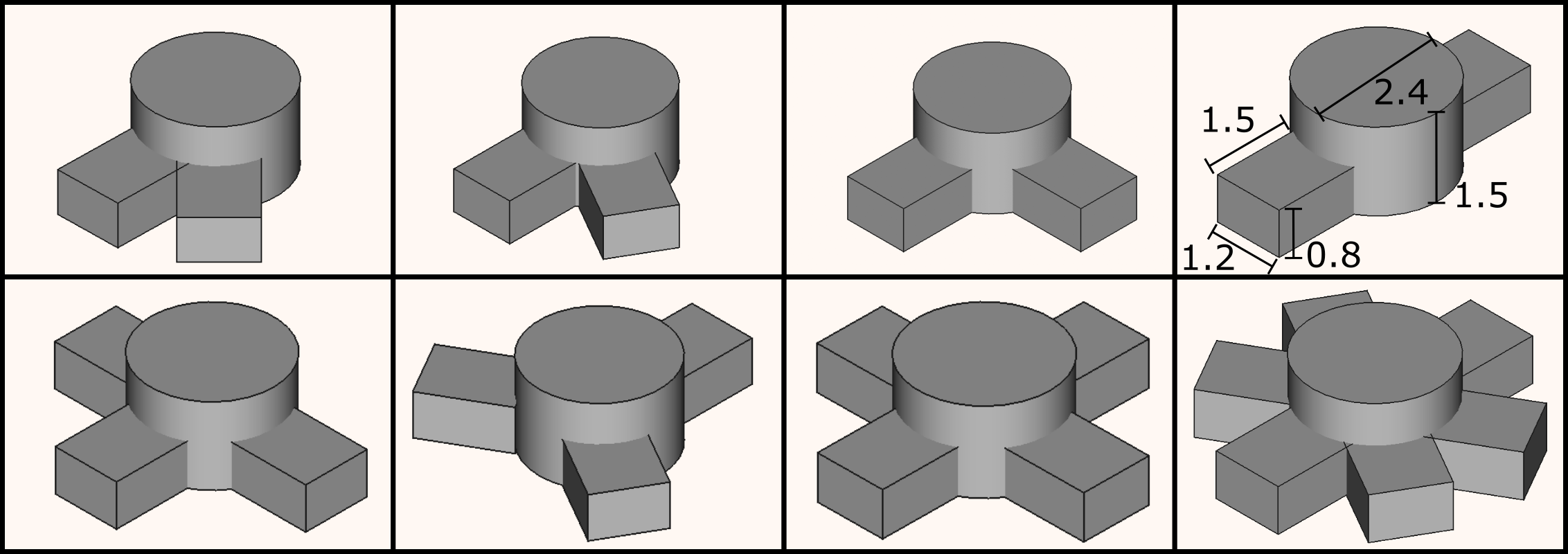}
  \caption{Schematic representation of the sCAPA traps for microgels designed with CAD. The dimensions are in \textmu m. }
  \label{fig:scapaMG}
\end{figure}

The final trap designs for the microgels where found by a parametric scan of of the different trap dimensions. The design consists of a cylinder that holds the 2 \textmu m PS-particle, surrounded by identical rectangular cuboids that hold the microgels. The dimensions of these fundamental shapes are constant for all the designs and can be found in Figure \ref{fig:scapaMG}.

\newpage

\subsection{Traps for printing-on-particles}
\begin{figure}[h]
  
   \centering
  
  \includegraphics[width=\linewidth]{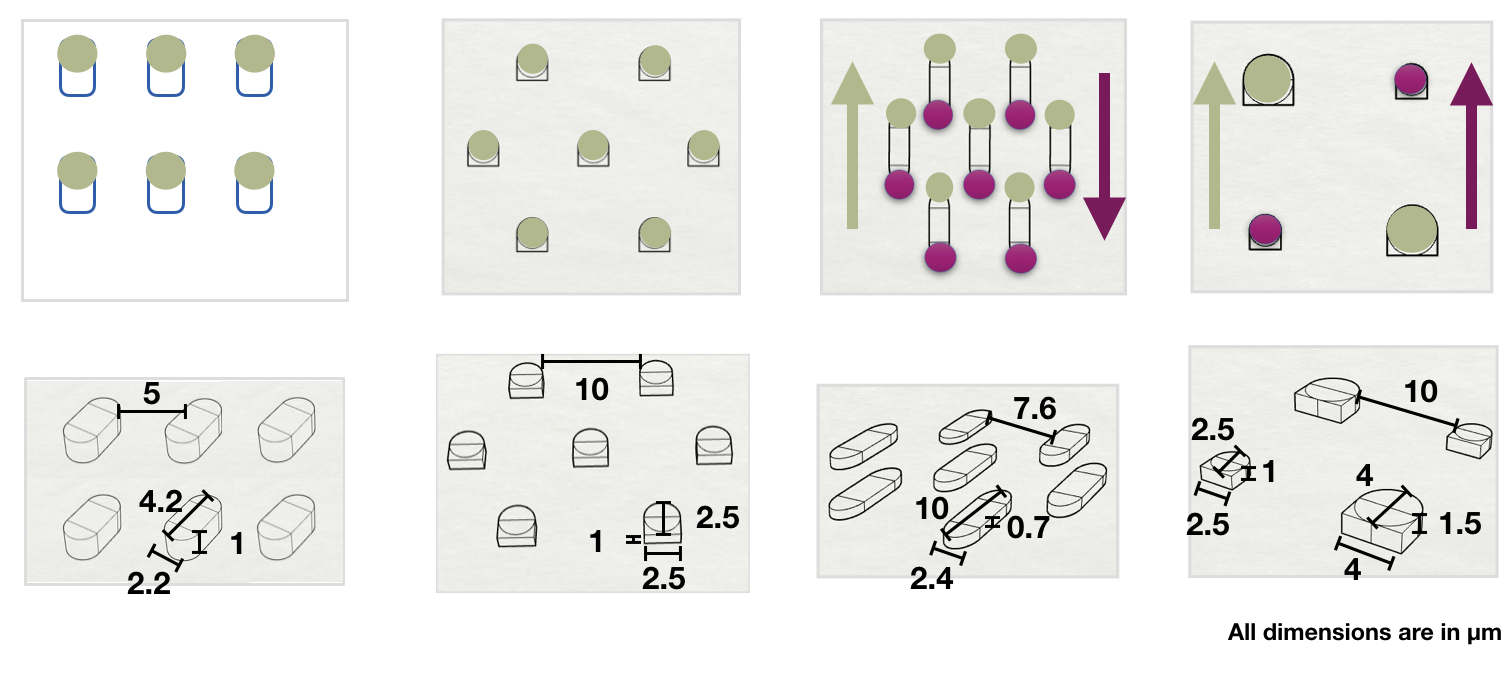}
  \caption{Schematic representation of the CAPA particle deposition process (top) and of the traps designed with CAD and the . The dimensions are in \textmu m. }
  \label{fig:scapa}
\end{figure}

\begin{figure}[h]
  \centering
  \includegraphics[width=0.8\linewidth]{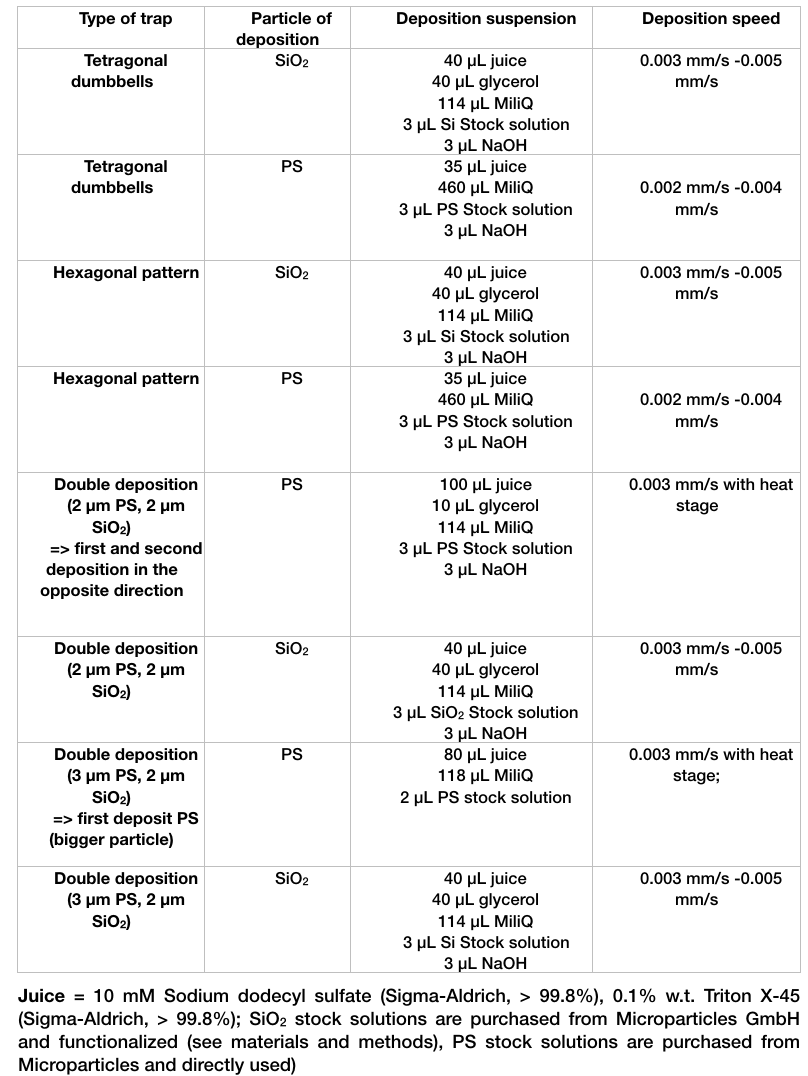}
   \caption{Table of parameters of the traps and working conditions of sCAPA deposition}
  \label{fig:scapa1}
\end{figure}

Traps of different sizes and shapes were designed with a CAD program (Rhinoceros 3D). We designed traps with round edges to better trap particles in the direction of deposition. sCAPA traps as well as the dimensions designed for each deposition are shown in Figure \ref{fig:scapa}, the working conditions are listed in Figure \ref{fig:scapa1}. The same trap parameters were used for depositing particles of the same size independent of the material. Longer sCAPA traps were designed for the double deposition of two particles of different materials of similar sizes, with 2 \textmu m silica deposited in one direction and the 2 \textmu m PS deposited in the opposite direction. Double deposition of two materials of different sizes were performed by combining single-particle traps of different sizes. Depositions of both particles were done in the same direction starting with the larger particle (3 \textmu m PS), followed by the smaller particle (2 \textmu m silica).
  
\clearpage

\section{sCAPA conditions}

\clearpage
\section{Tracking and DLW procedure}

The printing process in the DLW has the following steps. 

\begin{figure}[h]
  \centering
  \includegraphics[width=0.5\linewidth]{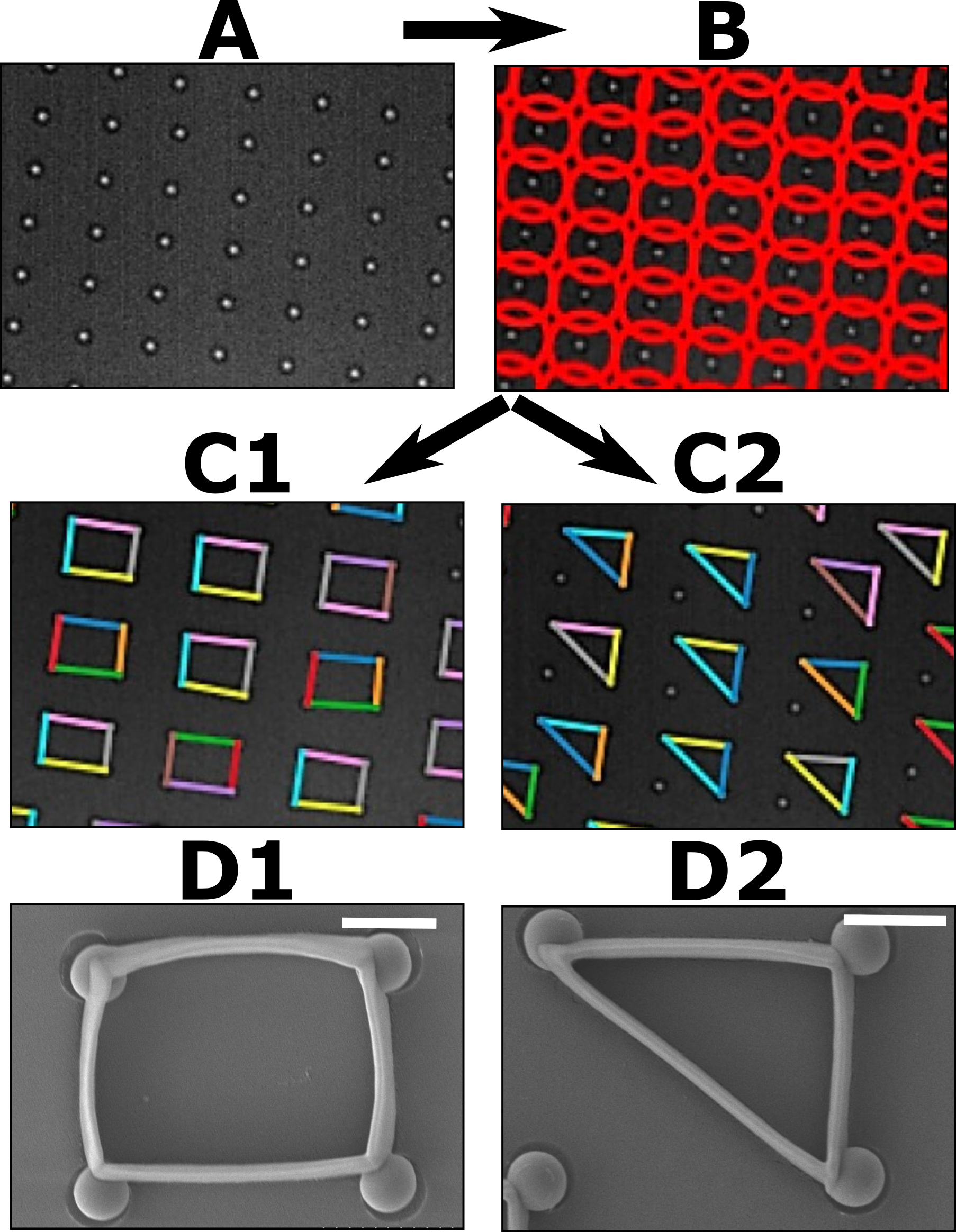}
  \caption{Tracking and 2PP DLW procedure. A) Reflection micrograph of the deposited particles covered in IP-L resin. B) Automated particle location by TrackPy, as marked by red circles. C) Visualization of the links to be printed, as defined by the user in Python. Examples for rectangles (C1) and triangles (C2) are shown. (D) SEM micrograph of 2 \textmu m SiO\textsubscript{2} particles deposited in a rectangular array and  linked in the shape of a rectangle (D1) and a triangle (D2) with IP-L after CPD. Scale bars (D1 and D2), 3 \textmu m.}
  \label{fig:fig3}
  
\end{figure}

\begin{itemize}
    \item Step 1. Take an image of the particles in the microscope (Fig. \ref{fig:fig3}A). The contrast is optimized by taking the image in reflection or transmission depending on the refractive index differences between the colloids and the resin.
    \item Step 2. Find the particles in the image (Fig. \ref{fig:fig3}A). This is done using TrackPy and all non-particles are filtered out by their minimum mass.
    \item Step 3. Find the nearest neighbours and categorize them by their relative angles. In the ideal scenario each colloid has 8 neighbours for the rectangular lattices and 6 neighbors for the hexagonal lattices. These are labeled to the cardinal directions (or A-F for the hexagons) and one can pick a sequence of labels, including the initial colloid (labeled "self"), to draw generate the lines between the colloids. For example, the sequence "Self,North,Northeast,East,Self" generates a rectangle (Fig. \ref{fig:fig3}C1) and "Self,East,Southeast,Self" generates a triangle (Fig. \ref{fig:fig3}C2).
    \item Step 4. The lines can be of different shapes which are generated from smaller line segments. These line segments are converted to printing coordinates by a pixel-to-\textmu m conversion with an offset based on manual calibrations.
    \item Step 5. The printing coordinates are written into a .GWL file that includes a header specifying the printing speed, printing power and offset in the z-direction. 
    \item Step 6. The print is carried out by the printer (Fig. \ref{fig:fig3}D1-2).
\end{itemize}


\clearpage

\section{Line printing: polymerization study}
\begin{figure}[h]
  \centering
  \includegraphics[width=\linewidth]{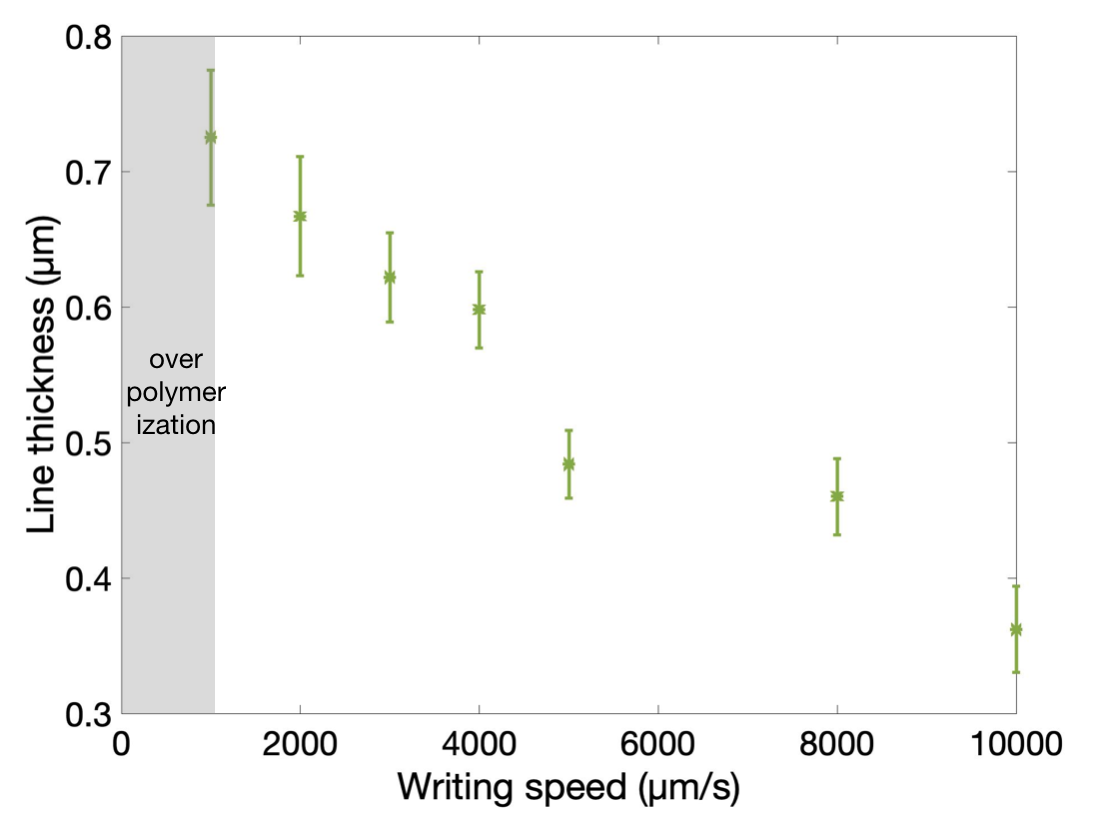}
  \caption{Plot of single-line thickness printed with IPL-resin as a function of printing speed.}
  \label{fig:chara1}
\end{figure}

\begin{figure}[h]
  \centering
  \includegraphics[width=\linewidth]{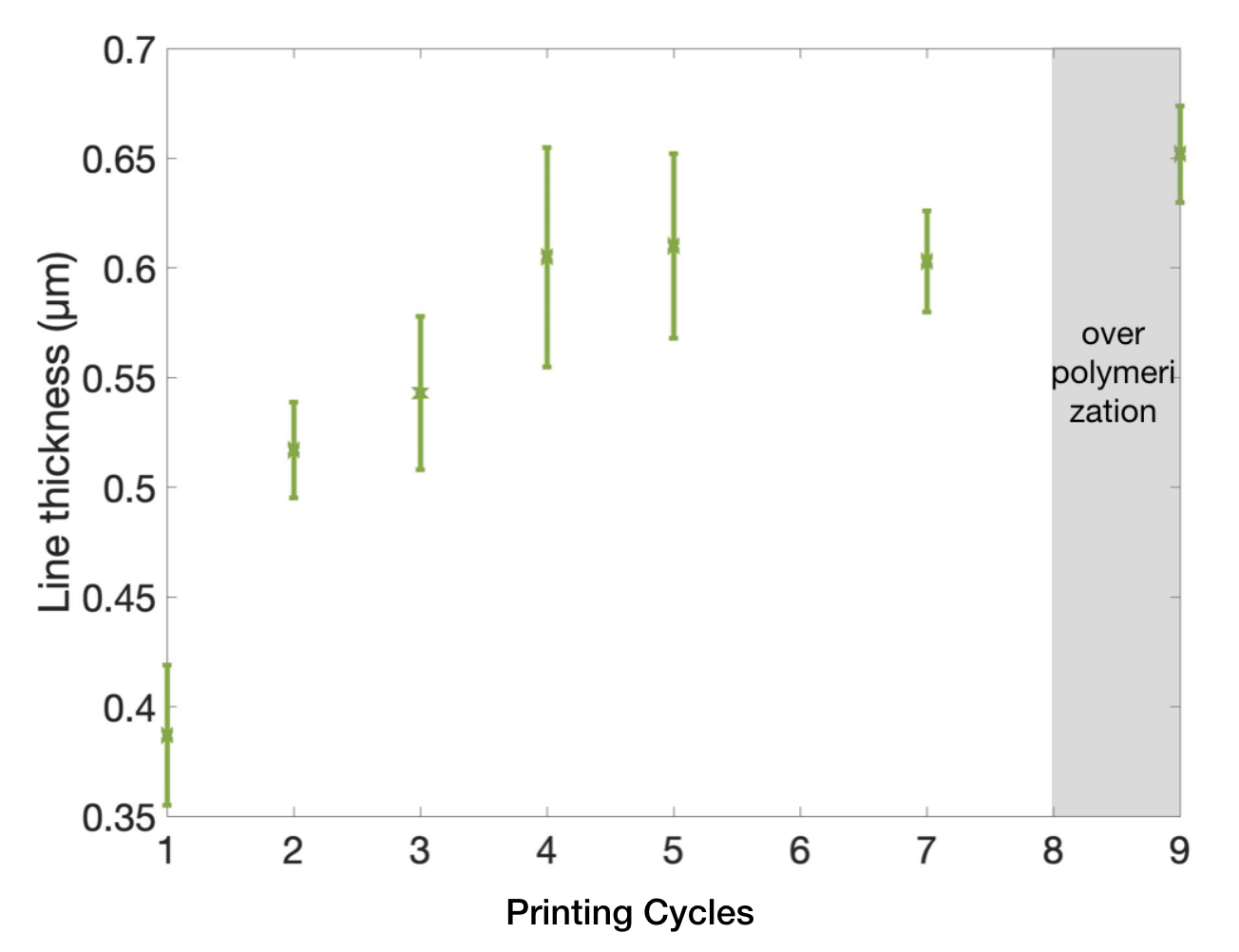}
  \caption{Plot of single-line thickness printed with IPL-resin as a function of the number of passages. With an increasing number of prints, the line thickness increases until it reaches a plateau of around 0.6 $\mu$m.}
  \label{fig:chara2}
\end{figure}

A parametric study of the printing conditions for IPL-resin is carried out to investigate the influence of scanning speed and printing times on resin polymerization, with the aim of measuring the thickness of single lines, which can be printed with IPL-resin by varying these parameters. Figure \ref{fig:chara1} shows a plot of line thickness versus laser scanning speed (writing speed). Single lines are printed three times at full laser power varying the writing speed. We observe that, with increasing laser scanning speed, the single-line thickness decreases. Overpolymerization occurs below 1000 $\mu$m/s resulting in bubble formation during printing. Figure \ref{fig:chara2} shows the plot of single line thickness versus the number of exposures (printing cycles) at constant scanning speed for the commercial IPL-resin. Single lines are printed at a constant speed of 10000 $\mu$m/s with full laser power. With increasing exposure cycles, the line thickness increases until a plateau of 0.6 $\mu$m is reached. Overpolymerization occurs with more than 8 cycles printing the same line.

\clearpage

\section{Additional 3D-printed colloidal molecules}
\begin{figure}[h]
  \centering
  \includegraphics[width=0.8\linewidth]{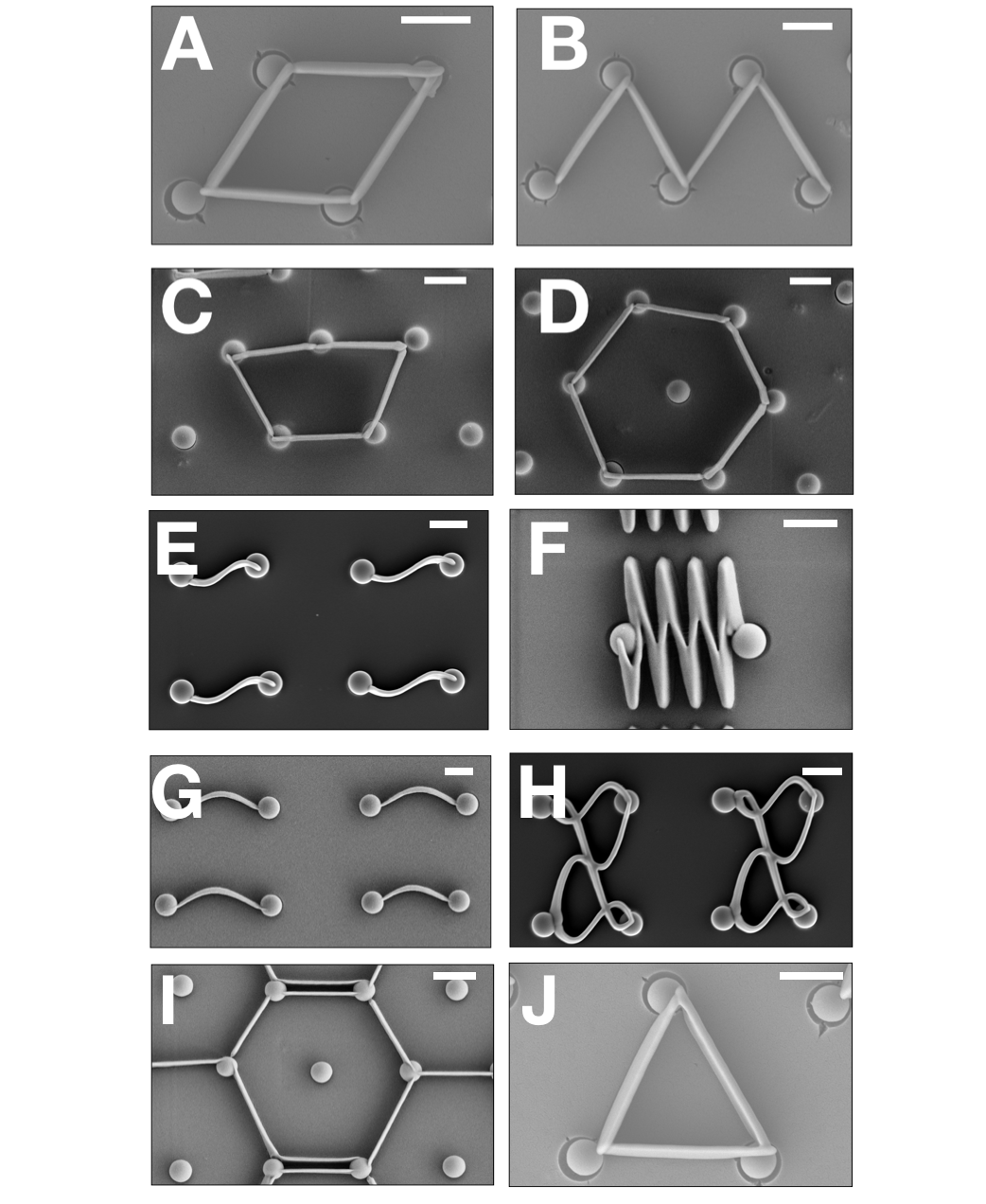}
  \caption{SEM images of neighboring CAPA deposited particles linked with IP-L in various shapes: Parallelogram (A), Letter M (B), Trapezoid (C), Hexagon (D), sine waves of multiple amplitudes and frequencies (E-G), colloidal calligraphy (H), colloidal 1,4-Cyclohexadiene (I) and equilateral triangle (J). The scale bars are 3 $\mu$m in length.}
\end{figure}

\newpage

\section{Release of a lattice}

\begin{figure} [h!]
  \centering
  \includegraphics[width=0.9\linewidth]{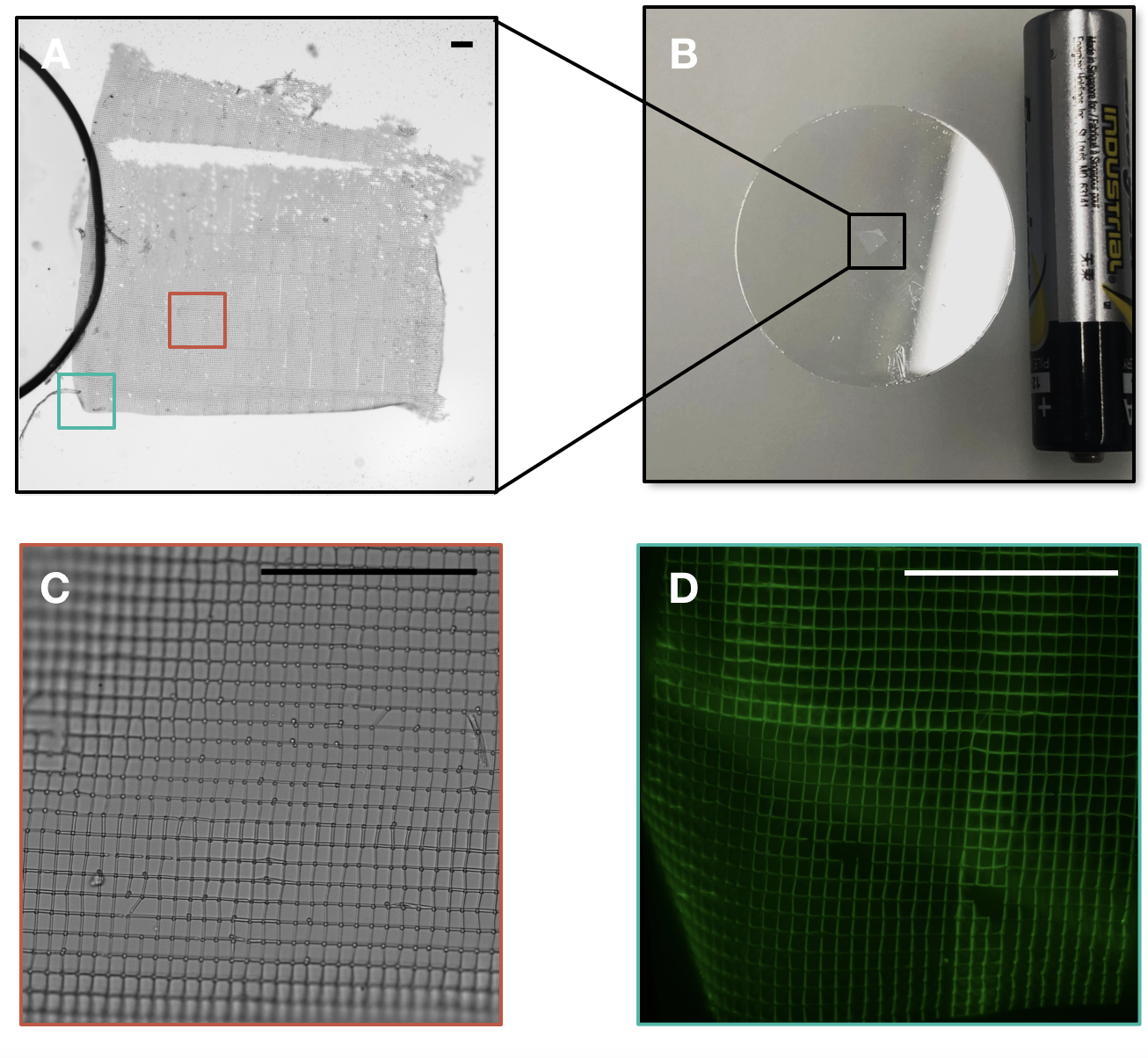}
  \caption{Micrograph of 2 \textmu m PS particles deposited on glucose-dextran sacrificial layer in the middle of the glass substrate and linked with IPL photoresist after developing and critical point drying (B). The size of the lattice is approx. 1.5 mm x 1.15 mm. Bright-field image of the linked lattice released in water (A). A zoomed in bright-field image of the lattice corresponding to the area in Figure A marked in red (C). A 3D fluorescent reconstruction of the edge of the released lattice corresponding to the area in Figure A marked in green (D). The scale bars are 100 \textmu m in length.}
  \label{fig:releaselattice}
\end{figure}

Figure \ref{fig:releaselattice} demonstrates the release of a large lattice of 2 \textmu m PS particles linked by IPL in water. 
The big lattice is composed of 170 smaller lattices (each lattice is one field of view) stitched together, the darker shades of prints are areas of stitching. They have been printed over twice with the same energy dosage. Figure \ref{fig:releaselattice} C is a zoomed in microscopic image of the whole lattice completely released and floating in water, with parts of the lattice in and out of focus. We observe that PS particles are incorporated inside the lattice.

